\documentclass[11pt]{article}

\usepackage{bbm}

\begin{document}

\renewcommand{\thefootnote}{\fnsymbol{footnote}}

\begin{titlepage}
\begin{center}
\hfill LTH 740\\
\hfill {\tt hep-th/0703036}\\
\vskip 10mm

{\Large
{\bf 
Duality and Black Hole Partition Functions\footnote{This
article is based on an a talk given at the 11th Marcel
Grossmann Meeting in Berlin, July 23 - 29, 2006.
}
}
}

\vskip 10mm

\textbf{Thomas Mohaupt}

\vskip 4mm

Theoretical Physics Division\\
Department of Mathematical Sciences\\
University of Liverpool\\
Liverpool L69 3BX, UK \\
{\tt Thomas.Mohaupt@liv.ac.uk}
\end{center}
 
\vskip .2in

\begin{center} {\bf ABSTRACT} \end{center}
\begin{quotation} \noindent
Supersymmetric black holes
provide an excellent theoretical laboratory to test ideas about quantum
gravity in general and black hole entropy in particular. When four-dimensional
supergravity is interpreted as the low-energy approximation of ten-dimensional
string theory or eleven-dimensional M-theory, one has a microscopic description
of the black hole which allows one to count microstates and to compare
the result to the macroscopic (geometrical) black hole entropy. Recently
it has been conjectured that there is a very direct connection between
the partition function of the topological string and a partition 
for supersymmetric black holes. We review this idea and propose 
a modification which makes it compatible with electric-magnetic duality.
\end{quotation}

\vfill

\end{titlepage}

\eject

\renewcommand{\thefootnote}{\arabic{footnote}}

\section{Supersymmetric Black Holes}

Our setup for constructing supersymmetric black hole solutions
is $N=2$ supergravity couled to $n$ vector multiplets. This 
arises\footnote{together
with further matter multiplets which are irrelevant for our purposes.}
as the effective field theory of heterotic string compactifications on 
$K3 \times T^2$ and of type-II string theory on Calabi-Yau threefolds.
The field equations are invariant under $Sp(2n+2,\mathbbm{R})$
rotations, which generalize the electric-magnetic duality rotations of
Einstein-Maxwell theory.\footnote{If the supergravity
action is the low energy effective action of a string compactification,
then string dualities, such as S-duality and T-duality, are embedded
into the symplectic group.}  As a consequence, all vector multiplet
couplings are encoded in a single holomorphic function called the
prepotential $F$. This function must be homogenous of degree 2 in 
its variables $Y^I$, which provide homogenous coordinates on the
scalar manifold $M_{\rm VM}$. The K\"ahler potential for the 
metric on $M_{\rm VM}$ can be expressed in terms of the holomorphic
prepotential. The resulting geometry is known as special (K\"ahler)
geometry \cite{deWit:1984pk,deWit:1984px}. It is possible to include a certain 
class of higher derivative terms involving the square of the Riemann 
tensor and arbitrary powers of gauge field strengths, 
by giving the prepotential an explicit dependence on the
so-called Weyl multiplet. 
%\cite{deWit:1996ix}. 
Associated to these
terms  is an infinite series of field-dependent couplings. In type-II
compactifications these couplings can be computed in terms of the
free energy of topologically twisted string theory 
\cite{Bershadsky:1993cx,Antoniadis:1993ze}.

%\section{Supersymmetric Black Holes}
%\label{sec:BPS_BHs}

As long as we neglect the higher derivative terms, we are dealing
with a generalized Einstein-Maxwell theory with several abelian
gauge fields and field-dependent couplings, plus a scalar
sigma-model. The supersymmetric black hole solutions of such a
theory are natural generalizations of the extremal Reissner-Nordstr\"om
black hole. Besides that the black hole now carries several electric
and magnetic charges, the new feature is that we have scalar fields
which vary non-trivially as a function of the radial variable.\footnote{
We only consider spherically symmetric solutions here.} At infinity, 
the solutions are asymptotically flat and the scalars can take
arbitrary values in $M_{\rm VM}$. The behaviour at the horizon 
is radically different: the scalars cannot take arbitrary values 
but must take fixed point values which are determined by the
electric and magnetic charges of the black hole. This is the so-called
black hole attractor mechanism \cite{Ferrara:1995ih}, which generalizes
to the case where higher derivative terms are included 
\cite{LopesCardoso:1998wt,LopesCardoso:2000qm}. Since both metric and
gauge fields are determined by the scalar fields through supersymmetry,
it follows that the area of the event horizon is a function of 
the electric and magnetic charges, and does not depend on the values
of the scalar fields at infinity. Once higher curvature
terms are included in the action, the black hole entropy is no
longer given by one quarter of the area of the event horizon\footnote{We
are using Planckian units.} but is given by the surface charge of
the Killing vector field which becomes null on the horizon \cite{Wald:1993nt}.
When evaluating the surface charge for supersymmetric black holes
in $N=2$ supergravity, one sees that the entropy is given by
the sum of two symplectic functions of the charges \cite{LopesCardoso:1998wt}.
While the first
term is the area of the horizon divided by 4, the second term depends
only on the couplings of the higher derivative terms. Therefore the 
black hole entropy is modified in two ways: first
through the modification of the area itself, second by the deviation
from the area law. The microscopic state 
degeneracy\cite{Maldacena:1997de,Vafa:1997gr} agrees with black hole
entropy if and only if both corrections are 
taken into account  \cite{LopesCardoso:1998wt}.

\section{Black Hole Partition Functions}

If one performs a partial Legendre transformation of the black hole
entropy,
which replaces the electric charges by the associated electrostatic
potentials, one obtains the imaginary part of the 
`generalized prepotential' \cite{Ooguri:2004zv}.
This is a power series in the Weyl multiplet which has 
as its coefficients the
prepotential (determining the two-derivative couplings) and
the coupling functions of the higher derivative terms.
By the relation between couplings in the effective
action and the topological string, this function is proportional
to the real part of the (holomorphic) free energy of the 
topologically twisted type-II string. This suggests to interprete
the imaginary part of the generalized prepotential as the 
free energy of the black hole, and one obtains the `OSV-relation' 
\cite{Ooguri:2004zv}
$Z_{\rm BH} = |Z_{\rm top}|^2$,
which relates the black hole partition function (exponential
of the free energy) to the partition function of the topological
string. However, many details of this proposal need to be made
more precise. One is whether the relation is meant to be an 
exact statement (strong version) or as
an asymptotic statement in the limit of large charges, which 
corresponds to the semi-classical limit (weak version).
Before reviewing the evidence supporting the weak version,
we need to address another point. By definition, the black hole free energy is
a function of the magnetic charges and of the electrostatic 
potential. Thermodynamically this corresponds to a mixed ensemble, 
where the magnetic charges have been fixed, while electric charges
fluctuate and the corresponding chemical potentials are fixed
\cite{Ooguri:2004zv}. This implies that a fundamental property,
namely covariance with respect to symplectic transformations is 
not manifest. As a consequence, it is not clear whether the proposal
is compatible with string dualities. In fact, discrepancies between
the actual microscopic state degeneracy and the state degeneracy
predicted by the OSV conjecture show that the OSV-relation must be
modified\cite{Shih:2005he,LopesCardoso:2006bg}.
%\footnote{These 
%calculations were
%performed for compactifications with $N=4$ supersymmetry, which, 
%however, can be treated within the $N=2$ formalism described in this
%article.} 
A natural way of deriving the modification is based on the
observation that the full Legendre transformation of the
black hole entropy, where both electric and magnetic charges
are replaced by the corresponding potentials 
has a natural meaning: the resulting
function is a Hesse potential for the metric on the scalar 
manifold \cite{LopesCardoso:2006bg}. Moreover, the relations between
entropy, free energy (mixed ensemble), Hesse potential and
attractor equations can be formulated in terms of a variational
principle \cite{Behrndt:1996jn,LopesCardoso:2006bg}. 
This suggests to interprete the Hesse potential 
as the free energy of the black hole, but now with respect to 
a canonical instead of a mixed ensemble. One can show that this 
proposal leads to a specific correction factor in the OSV-relation. 
Explicit tests can be performed in compactifications with $N=4$
supersymmetry, which can be treated within the $N=2$ formalism
explained in this article \cite{LopesCardoso:1999ur}. Subleading
corrections to the state degeneracy have been computed  
\cite{Dijkgraaf:1996it,LopesCardoso:2004xf,Jatkar:2005bh}
and the result agrees with the canonical black hole partition 
function proposed in \cite{LopesCardoso:2006bg} in the semi-classical
limit.
The agreement is impressive as it involves an infinite series of
non-perturbative corrections to the effective action.\footnote{These
corrections are world-sheet instantons from the point of view of
the type-II string but space-time instantons for the dual
heterotic string.}
The precise relation between the canonical black hole partition 
function and the topological string remains to be clarified. 
%Another open question is whether the OSV conjecture applies to
%so-called small black holes,\footnote{These are solution which 
%have a null singularity, which is resolved by higher derivative
%terms.} where the subleading corrections to macroscopic and
%microscopic entropy do not quite agree. 

%\bibliographystyle{ws-procs975x65}
%\bibliography{LitMG11}

\begin{thebibliography}{10}

\bibitem{deWit:1984pk}
B.~de~Wit and A.~Van~Proeyen, {\em Nucl. Phys.} {\bf B245}, p.~89 (1984).

\bibitem{deWit:1984px}
B.~de~Wit, P.~G. Lauwers and A.~Van~Proeyen, {\em Nucl. Phys.} {\bf B255}, p.
  569 (1985).

\bibitem{Bershadsky:1993cx}
M.~Bershadsky, S.~Cecotti, H.~Ooguri and C.~Vafa, {\em Commun. Math. Phys.}
  {\bf 165}, 311 (1994).

\bibitem{Antoniadis:1993ze}
I.~Antoniadis, E.~Gava, K.~S. Narain and T.~R. Taylor, {\em Nucl. Phys.} {\bf
  B413}, 162 (1994).

\bibitem{Ferrara:1995ih}
S.~Ferrara, R.~Kallosh and A.~Strominger, {\em Phys. Rev.} {\bf D52}, 5412
  (1995).

\bibitem{LopesCardoso:1998wt}
G.~Lopes~Cardoso, B.~de~Wit and T.~Mohaupt, {\em Phys. Lett.} {\bf B451}, 309
  (1999).

\bibitem{LopesCardoso:2000qm}
G.~Lopes~Cardoso, B.~de~Wit, J.~Kappeli and T.~Mohaupt, {\em JHEP} {\bf 12}, p.
  019 (2000).

\bibitem{Wald:1993nt}
R.~M. Wald, {\em Phys. Rev.} {\bf D48}, 3427 (1993).

\bibitem{Maldacena:1997de}
J.~M. Maldacena, A.~Strominger and E.~Witten, {\em JHEP} {\bf 12}, p. 002
  (1997).

\bibitem{Vafa:1997gr}
C.~Vafa, {\em Adv. Theor. Math. Phys.} {\bf 2}, 207 (1998).

\bibitem{Ooguri:2004zv}
H.~Ooguri, A.~Strominger and C.~Vafa, {\em Phys. Rev.} {\bf D70}, p. 106007
  (2004).

\bibitem{Shih:2005he}
D.~Shih and X.~Yin, {\em JHEP} {\bf 04}, p. 034 (2006).

\bibitem{LopesCardoso:2006bg}
G.~Lopes~Cardoso, B.~de~Wit, J.~Kappeli and T.~Mohaupt, {\em JHEP} {\bf 03}, p.
  074 (2006).

\bibitem{Behrndt:1996jn}
K.~Behrndt {\em et~al.}, {\em Nucl. Phys.} {\bf B488}, 236 (1997).

\bibitem{LopesCardoso:1999ur}
G.~Lopes~Cardoso, B.~de~Wit and T.~Mohaupt, {\em Nucl. Phys.} {\bf B567}, 87
  (2000).

\bibitem{Dijkgraaf:1996it}
R.~Dijkgraaf, E.~P. Verlinde and H.~L. Verlinde, {\em Nucl. Phys.} {\bf B484},
  543 (1997).

\bibitem{LopesCardoso:2004xf}
G.~Lopes~Cardoso, B.~de~Wit, J.~Kappeli and T.~Mohaupt, {\em JHEP} {\bf 12}, p.
  075 (2004).

\bibitem{Jatkar:2005bh}
D.~P. Jatkar and A.~Sen, {\em JHEP} {\bf 04}, p. 018 (2006).

\end{thebibliography}

\end{document}